\documentclass[preprint,eqsecnum,preprintnumbers,nofootinbib,byrevtex,prd,aps,showpacs,showkeys,groupedaddress,floatfix]{revtex4}
\usepackage{bm}
\usepackage{graphics}
\usepackage{graphicx}
\usepackage{epsfig}
\usepackage{amssymb}
\usepackage{amsmath}
\begin{document}

%\date{}
%\large
\title{Any l-state analytical solutions of the  Klein-Gordon equation for the Woods-Saxon potential}
\author{V.~H.~Badalov$^{1}$} \email{E-mail:badalovvatan@yahoo.com}
\author{H.~I.~Ahmadov$^{2}$}
\author{S.~V.~Badalov$^{3}$}%
\affiliation{$^{1}$Institute for Physical Problems, 
Baku State University, Z. Khalilov st. 23, AZ-1148 Baku,
Azerbaijan\\
 $^{2}$\ Department of Mathematical Physics,Faculty of
Applied Mathematics and Cybernetics,Baku State University, Z.
Khalilov st. 23, AZ-1148 Baku, Azerbaijan\\
$^{3}$\ Department of Physics, Bilkent University, TR-06800 Ankara,
Turkey}

%\date{}

\begin{abstract}
{The radial part of the Klein-Gordon equation for the Woods-Saxon
potential is solved. In our calculations, we have applied the
Nikiforov-Uvarov method  by using the Pekeris approximation to the
centrifugal potential for any $l$ states. The exact bound state
energy eigenvalues and the corresponding eigenfunctions are
obtained on the various values of the quantum numbers $n$ and $l$.
The non-relativistic limit of the bound state energy spectrum was
also found.}
\end{abstract}


\maketitle

\section{\bf Introduction}

An analytical solution of the radial part of the Klein-Gordon
equation is of high importance spinless in relativistic quantum
mechanics, because the wave function contains all necessary
information for full description of a quantum system. There are
only few potentials for which the radial part of the Klein-Gordon
equation can be solved explicitly for all $n$ and $l$. So far,
many methods were developed, such as supersymmetry (SUSY) [1,2]
and the Pekeris approximation [3-8], to solve radial part of the
Klein-Gordon equation exactly or quasi-exactly for $l\neq0$ within
these potentials.

The one-dimensional Klein-Gordon (KG) equation is investigated for
the PT -symmetric generalized Woods-Saxon (WS) potential [9] and
Hulth\'en [10] and is solved by using the Nikiforov-Uvarov (NU)
method which is based on solving the second-order linear
differential equations by reduction to a generalized equation of
hypergeometric type.

The radial part of the Klein-Gordon equation for the Woods-Saxon
potential [11] cannot be solved exactly for $l\neq0$. But in
Refs.[9, 12-14] authors as in Refs.[15, 16] because of errors made
by them in application of the NU method to investigate the
Woods-Saxon potential obtained wrong results. It is well known that
the Woods-Saxon potential is one of the important short-range
potentials in physics. And this potential was applied to numerous
problems, in nuclear and particle physics, atomic physics, condensed
matter and chemical physics. Therefore, it would be interesting and
important to solve the radial of the Klein - Gordon equation for
Woods-Saxon potential for $l\neq0$, since it has been extensively
used to describe the bound and continuum states of the interacting
systems.

In this work, we solve  the radial part of the Klein-Gordon
equation for the standard Woods-Saxon potential using NU method
[17], and obtain the energy eigenvalues and corresponding
eigenfunctions for any $l$ states.

\section{\bf Nikiforov-Uvarov method}

The Nikiforov-Uvarov (NU) method is based on the solutions of
general second-order linear equations with special orthogonal
functions. It has been extensively used to solve the nonrelativistic
Schr\"{o}dinger equation and other Schr\"{o}dinger-like equations.
The one-dimensional Schr\"{o}dinger equation or similar second-order
differential equations can be written using NU method in the
following form:
\begin{equation}
\psi''(z)+\frac{\widetilde{\tau}(z)}{\sigma(z)}{\psi}'(z)+
\frac{\widetilde{\sigma}(z)}{\sigma^{2}(z)}\psi(z)=0,
\end{equation}
where $\sigma(z)$ and $\widetilde{\sigma}(z)$ are polynomials, at
most second-degree, and $\widetilde{\tau}(z)$ is a first-degree
polynomial.

Using Eq(2.1) the transformation
\begin{equation}
\psi(z)=\Phi(z)\\{y}(z)
\end{equation}
one reduces it to the hypergeometric-type equation
\begin{equation}
\sigma(z){y}''+\tau(z){y}'+\lambda{y}=0.
\end{equation}
The function $\Phi(z)$ is defined as the logarithmic derivative [10]
\begin{equation}
\frac{\Phi'(z)}{\Phi(z)}=\frac{\pi(z)}{\sigma(z)},
\end{equation}
where $\pi(z)$ is at most the first-degree polynomial.

The another part of $\psi(z)$, namely ${y}(z)$, is the
hypergeometric-type function, that for fixed $n$ is given by the
Rodriguez relation:
\begin{equation}
{y_{n}}(z)=\frac{{B_{n}}}{\rho(z)}\frac{{d^{n}}}{{dz^{n}}}[\sigma^{n}(z)\rho(z)],
\end{equation}
where ${B_{n}}$ is the normalization constant and the weight
function $\rho(z)$ must satisfy the condition [12]
\begin{equation}
\frac{d}{dz}\left(\sigma(z)\rho(z)\right)=\tau(z)\rho(z),
\end{equation}
with $\tau(z)=\widetilde{\tau}(z)+2\pi(z).$

For accomplishment of the conditions imposed on function $\rho(z)$
of the classical orthogonal polynomials, it is necessary, that
polynomial $\tau(z)$ becomes equal to zero in some point of an
interval $(a,b)$ and derivative of this polynomial for this interval
at $\sigma(z)>0$ will be negative, i.e., $\tau'(z)<0$.

The function $\pi(z)$ and the parameter $\lambda$ required for this
method are defined as follows:
\begin{equation}
\pi(z)=\frac{\sigma'-\widetilde{\tau}}{2}\pm\sqrt{\left(\frac{\sigma'-
\widetilde{\tau}}{2}\right)^{2}-\widetilde{\sigma}+{k}\sigma},
\end{equation}
\begin{equation}
\lambda=k+\pi'(z).
\end{equation}
On the other hand, in order to find the value of $k$, the expression
under the square root must be the square of a polynomial. This is
possible only if its discriminant is zero. Thus, the new eigenvalue
equation for the Eq.(2.1) is [17]:
\begin{equation}
\lambda=\lambda_{n}=-n\tau'-\frac{n(n-1)}{2}\sigma'',\,\,
(n=0,1,2,...),
\end{equation}
After the comparison of Eq.(2.8) with Eq.(2.9), we obtain the energy
eigenvalues.

\section{\bf Solutions of the Klein-Gordon equation with the
Woods-Saxon potential}

The standard Woods-Saxon potential [11] is defined by
\begin{equation}
V(r)=-\frac{V_{0}}{1+\exp\left(\frac{r-R_{0}}{a}\right)},\,\,\, a\ll
R_{0},\,\,\, 0\leq r<\infty.
\end{equation}

This potential was used for description of interaction of a neutron
with a heavy nucleus. The parameter $R_{0}$ is interpreted as radius
of a nucleus, the parameter $a$ characterizes thickness of the
superficial layer inside, which the potential falls from value $V=0$
outside of a nucleus up to value $V=-V_{0}$ inside of a nucleus. At
$a=0$, one gets the simple potential well with jump of potential on
the surface of a nucleus.

In the spherical coordinates, the stationary Klein-Gordon equation
with Woods-Saxon potential is [18]

 \begin{equation}
-\hbar^{2}c^{2}\left\{\frac1{r^{2}}\frac\partial{\partial
r}\left(r^{2}\frac\partial{\partial r}\right)+\left[\frac1{r^{2}\sin
\theta}\frac\partial{\partial\theta}\left(\sin\theta\frac
\partial
{\partial\theta}\right)+\frac1{r^{2}\sin^{2}\theta}\frac{\partial^{2}}{
\partial\varphi^{2}}\right]\right\}\psi=
\left[\left(E-V(r)\right)^{2}-m_{0}^{2}c^{4}\right]\psi.
\end{equation}
where $m_{0}$ is the rest mass of a scalar particle.

The terms in the square brackets with the overall minus sign are
the dimensionless angular momentum squared operator,
$\widehat{L}^{2}$. Defining $\psi(r,\theta,\varphi
)=R(r)Y(\theta,\varphi)$, we obtain the radial part of the
Klein-Gordon equation [18]
\begin{equation}
\frac{d^{2}R(r)}{dr^{2}}+\frac
2r\frac{dR(r)}{dr}+\left[\frac{\left(E-V\right)^{2}-m_{0}^{2}c^{4}}{\hbar^{2}c^{2}}-\frac{l\left(
l+1\right)}{r^{2}}\right]R(r)=0.
\end{equation}
where $l$ is the angular momentum quantum number.

After introducing the new function: $$u(r)=rR(r),$$ Eq.(3.3) takes
the form:

\begin{equation}
\frac{d^{2}u(r)}{dr^{2}}+\left[\frac{\left(E-V\right)^{2}-m_{0}^{2}c^{4}}{\hbar^{2}c^{2}}-\frac{l\left(
l+1\right)}{r^{2}}\right]u(r)=0.
\end{equation}

The Eq.(3.4) has the same form as the equation for a particle in one
dimension, except for two important differences. First, there is a
repulsive effective potential proportional to the eigenvalue of
$\hbar^{2}l(l+1)$. Second, the radial function must satisfy the
boundary conditions $u(0)=0$ and $u(\infty)=0.$

If in Eq.(3.4) introduce the notations:
$$x=\frac{r-R_{0}}{R_{0}},\,\,\, \alpha=\frac{R_{o}}{a},$$
then the Woods-Saxon potential is given by the expression:
$$V_{WS}=-\frac{V_{0}}{1+\exp(\alpha x)}.$$

It is known that the radial part Klein-Gordon equation cannot be
solved exactly for this potential at the value $l\neq0$ using the
standard methods as SUSY and NU. From Eq.(3.4) it is seen that the
centrifugal potential
$V_{l}(r)=\frac{\hbar^{2}l(l+1)}{2m_{0}r^{2}}$ is inverse square
potentials and as a result, this equation (3.4) cannot be solved
analytically. Therefore, in order to solve this problem we can
take the most widely used and convenient for our purposes Pekeris
approximation. This approximation is based on the expansion of the
centrifugal barrier in a series of exponentials depending on the
internuclear distance, taking into account terms up to second
order, so that the effective $l$ dependent potential preserves the
original form. It should be pointed out, however, that this
approximation is valid only for low vibrational energy states. By
changing the coordinates $x=\frac{r-R_{0}}{R_{0}}$ or
$r=R_{0}(1+x)$, the centrifugal potential is expanded in the
Taylor series around the point $x=0$ ($r=R_{0}$)
\begin{equation}
V_{l}(r)=\frac{\hbar^{2}l(l+1)}{2m_{0}r^{2}}=\frac{\hbar^{2}l(l+1)}
{2m_{0}R_{0}^{2}}\frac{1}{(1+x)^{2}}=\delta\left(1-2x+3x^{2}-4x^{3}+\ldots\right),
\end{equation}
where $\delta=\frac{\hbar^{2}l(l+1)}{2m_{0}R_{0}^{2}}$.

According to the Pekeris approximation, we shall replace potential
$V_{l}(r)$ with expression [7,8]:

\begin{equation}
V^{*}_{l}(r)=\delta \left(C_{0}+\frac{C_{1}}{1+\exp\alpha
x}+\frac{C_{2}}{\left(1+\exp\alpha x\right)^{2}}\right),
\end{equation}
where $C_{0}$, $C_{1}$ and $C_{2}$ are some constants.

In order to define the constants $C_{0}$, $C_{1}$ and $C_{2}$, we
also expand this potential in the Taylor series around the point
$x=0$ ($r=R_{0}$):
\begin{equation}
V^{*}_{l}(x)=\delta\left[\left(C_{0}+\frac{C_{1}}{2}+\frac{C_{2}}{4}\right)-
\frac{\alpha}{4}\left(C_{1}+C_{2}\right)x+\frac{\alpha^{2}}{16}C_{2}x^{2}+\frac{\alpha^{3}
}{48}\left(C_{1}+C_{2}\right)x^{3}-\frac{\alpha^{4}}{96}C_{2}x^{4}+\cdots\right].
\end{equation}

Comparing equal powers of $x$ Eqs.(3.5) and (3.7), we obtain the
constants $C_{0}$, $C_{1}$ and $C_{2}$ [7,8]:
$$C_{0}=1-\frac{4}{\alpha}+\frac{12}{\alpha^{2}},\,\,\,
C_{1}=\frac{8}{\alpha}-\frac{48}{\alpha^{2}},\,\,\,
C_{2}=\frac{48}{\alpha^{2}}.$$

Instead of solving the radial part Klein-Gordon equation for the
centrifugal potential $V_{l}(r)$  given by Eq.(3.5), we now solve
the radial part Klein-Gordon equation for the new centrifugal
potential $V^{*}_{l}(r)$ given by Eq.(3.6) obtained using the
Pekeris approximation. Having inserted this new centrifugal
potential into Eq.(3.4), we obtain:
\begin{equation}
\frac{d^2u}{dr^2}+\left[\left(\frac{E^{2}-m_{0}^{2}c^{4}}{\hbar^{2}c^{2}}-\frac{l\left(
l+1\right)C_{0}}{\alpha^{2}a^{2}}\right)+\frac{\left(\frac{2EV_{0}}{\hbar^{2}c^{2}}-\frac{l\left(
l+1\right)C_{1}}{\alpha^{2}a^{2}}\right)}{{1+e^{\frac{r-R_{0}}{a}}}}+\frac{\left(\frac{V_{0}^{2}}{\hbar^{2}c^{2}}-\frac{l\left(
l+1\right)C_{2}}{\alpha^{2}a^{2}}\right)}{\left(1+e^{\frac{r-R_{0}}{a}}\right)^{2}}\right]u=0.
\end{equation}
We use the following dimensionless notations:
\begin{equation}
\epsilon^{2}=-\left(\frac{\left(E^{2}-m_{0}^{2}c^{4}\right)a^{2}}{\hbar^{2}c^{2}}-\frac{l\left(
l+1\right)C_{0}}{\alpha^{2}}\right);\,\,\beta^2=\frac{2EV_{0}a^{2}}{\hbar^{2}c^{2}}-\frac{l\left(
l+1\right)C_{1}}{\alpha^{2}};\,\,
\gamma^{2}=-\left(\frac{V_{0}^{2}a^{2}}{\hbar^2c^{2}}-\frac{l\left(
l+1\right)C_{2}}{\alpha^{2}}\right),
\end{equation}
with  real $\epsilon>0$ $(E^{2}\leq m_{0}^{2}c^{2})$ for bound
states; $\beta$ is real, $\gamma$ is real and positive.

If we rewrite Eq.(3.8) by using a new variable of the form [4]:
$$z=\left(1+\exp\left(\frac{r-R_{0}}{a}\right)\right)^{-1},$$
we obtain:
\begin{equation}
u^{\prime \prime }(z)+\frac{1-2z}{z(1-z)}u^{\prime}(z)+
\frac{-\epsilon^{2}+\beta^{2}z-\gamma^{2}z^{2}}{(z(1-z))^{2}}u(z)=0,
(0\leq z\leq 1),
\end{equation}
with $\widetilde{\tau}(z)=1-2z;\,\,\, \sigma (z)=z(1-z);\,\,\,
\widetilde{\sigma}(z)=-\epsilon^2+\beta^{2}z-\gamma^{2}z^{2}$.

In the NU-method the new function $\pi(z)$ is:
\begin{equation}
\pi(z)=\pm\sqrt{\epsilon^{2}+\left(k-\beta^{2}\right)z-\left(k-\gamma^{2}\right)z^{2}}.
\end{equation}

The constant parameter $k$ can be found employing the condition
that the expression under the square root has a double zero, i.e.,
its discriminant is equal to zero. So, there are two possible
functions for each $k$:
\begin{equation}
\label{2}\pi(z)=\pm\left\{
\begin{array}{c}
\left(\epsilon-\sqrt{\epsilon^{2}-\beta^{2}+\gamma^{2}}\right)z-\epsilon,\,\,\,
for \,\,\,
k=\beta^{2}-2\epsilon^{2}+2\epsilon\sqrt{\epsilon^{2}-\beta^{2}+\gamma^{2}},\\
\left(\epsilon+\sqrt{\epsilon^{2}-\beta^{2}+\gamma^{2}}\right)z-\epsilon,\,\,\,
for \,\,\,\,
k=\beta^{2}-2\epsilon^{2}-2\epsilon\sqrt{\epsilon^{2}-\beta^{2}+\gamma^{2}}.\,
\end{array}
\right.
\end{equation}

According to the NU-method, from the four possible forms of the
polynomial $\pi(z)$ we select the one for which the function
$\tau(z)$ has the negative derivative and root lies in the interval
(0,1). Therefore, the appropriate functions $\pi(z)$ and $\tau(z)$
have the following forms:
\begin{equation}
\pi(z)=\epsilon-\left(\epsilon+\sqrt{\epsilon^{2}-\beta^{2}+\gamma^{2}}\right)z,
\end{equation}

\begin{equation}
\tau(z)=1+2\epsilon-2\left(1+\epsilon+\sqrt{\epsilon^{2}-\beta^{2}+\gamma^{2}}\right)z,
\end{equation}
and
\begin{equation}
k=\beta^{2}-2\epsilon^{2}-2\epsilon\sqrt{\epsilon^{2}-\beta^{2}+\gamma^{2}}.
\end{equation}
Then, the constant $\lambda=k+\pi'(z)$ is written as:
\begin{equation}
\lambda=\beta^{2}-2\epsilon^{2}-2\epsilon\sqrt{\epsilon^{2}-\beta^{2}+\gamma^{2}}-
\epsilon-\sqrt{\epsilon^{2}-\beta^{2}+\gamma^{2}}.
\end{equation}
An alternative definition of $\lambda_{n}$ (Eq.(2.9)) is:
\begin{equation}
\lambda=\lambda_{n}=2\left(\epsilon+\sqrt{\epsilon^{2}-
\beta^{2}+\gamma^{2}}\right)n+n(n+1).
\end{equation}
Having compared Eqs.(3.16) and (3.17):
\begin{equation}
\beta^{2}-2\epsilon^{2}-2\epsilon\sqrt{\epsilon^{2}-\beta^{2}+\gamma^{2}}-
\epsilon-\sqrt{\epsilon^{2}-\beta^{2}+\gamma^{2}}=2\left(\epsilon+\sqrt{\epsilon^{2}-
\beta^{2}+\gamma^{2}}\right)n+n(n+1),
\end{equation}
we obtain
\begin{equation}
\epsilon+\sqrt{\epsilon^{2}-\beta^{2}+\gamma^{2}}+
n+\frac{1}{2}-\frac{\sqrt{1+4\gamma^{2}}}{2}=0,
\end{equation}
or
\begin{equation}
\epsilon+\sqrt{\epsilon^{2}-\beta^{2}+\gamma^{2}}-n'=0.
\end{equation}
Here
\begin{equation}
n'=-n+\frac{\sqrt{1+4\gamma^{2}}-1}{2},
\end{equation}
$n$ being the radial quantum number $(n=0,1,2,\ldots)$.

After substituting $\alpha, \gamma, C_{2}$ into Eq.(3.21), we
obtain:

\begin{equation}
n'=-n+\frac{\sqrt{1+\frac{192a^4l(l+1)}
{R_{0}^4}-\frac{4V_{0}^{2}a^{2}}{\hbar ^{2}c^{2}}}-1}{2}.
\end{equation}

From Eq.(3.20), we find:

\begin{equation}
\epsilon=\frac{1}{2}\left(n'+\frac{\beta^{2}-\gamma^{2}}{n'}\right).
\end{equation}

Because for the bound states $\epsilon>0$, we get:

\begin{equation}
n'>0.
\end{equation}

 This physically interesting point imposes that $n'>0$, which certainly
determines the number of physically meaningful bound states for a
deep potential appearing near the surface. If $n'>0$, there exist
bound states, otherwise, there are no bound states at all. By using
Eq.(3.22) this relation can be recast into the form:

\begin{equation}
0\leq n<\frac{\sqrt{1+\frac{192l(l+1)a^{4}}
{R_{0}^{4}}-\frac{4V_{0}^{2}a^{2}}{\hbar ^{2}c^{2}}}-1}{2}
\end{equation}
i.e. it gives the finite coupling value.

The condition $\gamma>0$ gives the define coupling value for the
potential depth $V_{0}$:
\begin{equation} 0<V_{0}<\frac{4\hbar
ca\sqrt{3l(l+1)}}{R_{0}^{2}}.
\end{equation}

From Eqs. (3.23) and (3.9), we obtain:

\begin{equation}
\begin{array}{c}
E_{nl}=-\frac{V_{0}}{2}\left(1-\frac{l(l+1)(C_{1}+C_{2})}{\alpha
^{2}\left(n'^{2}+\frac{4V_{0}^{2}a^{2}}{\hbar
^{2}c^{2}}\right)}\right) \pm
cn'\sqrt{\frac{m_{0}^{2}c^{2}+\frac{\hbar
^{2}l(l+1)}{R_{0}^{2}}C_{0}}{n'^{2}+\frac{4V_{0}^{2}a^{2}}{\hbar
^{2}c^{2}}}-\frac{\hbar
^{2}}{4a^{2}}\left(1-\frac{l(l+1)(C_{1}+C_{2})}{\alpha
^{2}\left(n'^{2}+\frac{4V_{0}^{2}a^{2}}{\hbar
^{2}c^{2}}\right)}\right)^{2}}
\end{array}
\end{equation}

Thus, substituting the expressions of $\alpha, C_{0}, C_{1}, C_{2}$
and $n'$ into Eq.(3.27), one can find the energy eigenvalues
$E_{nl}$
\begin{equation}
\begin{array}{c}
E_{nl}=-\frac{V_{0}}{2}\left(1-\frac{32l(l+1)a^{3}}{R_{0}^{3}\left[\left(
\sqrt{1+\frac{192l(l+1)a^{4}}{R_{0}^{4}}-\frac{4V_{0}^{2}a^{2}}{\hbar
^{2}c^{2}}}-2n-1\right)^{2}+\frac{4V_{0}^{2}a^{2}}{\hbar
^{2}c^{2}}\right]}\right) \left. \right. \pm    \\
c\left(
\sqrt{1+\frac{192l(l+1)a^4}{R_{0}^{4}}-\frac{4V_{0}^{2}a^{2}}{\hbar
^{2}c^{2}}}-2n-1\right)\left\{\frac{m_{0}^{2}c^{2}+\frac{\hbar
^{2}l(l+1)}{R_{0}^{2}}\left(1-\frac{4a}{R_{0}}+
\frac{12a^{2}}{R_{0}^{2}}\right)}{\left(
\sqrt{1+\frac{192l(l+1)a^{4}}{R_{0}^{4}}-\frac{4V_{0}^{2}a^{2}}{\hbar
^{2}c^{2}}}-2n-1\right)^{2}+\frac{4V_{0}^{2}a^{2}}{\hbar
^{2}c^{2}}}-   \right.   \\   \left.  \frac{\hbar
^{2}}{16a^{2}}\left(1-\frac{32l(l+1)a^{3}}{R_{0}^{3}\left[\left(
\sqrt{1+\frac{192l(l+1)a^{4}}{R_{0}^{4}}-\frac{4V_{0}^{2}a^{2}}{\hbar
^{2}c^{2}}}-2n-1\right)^{2}+\frac{4V_{0}^{2}a^{2}}{\hbar
^{2}c^{2}}\right]}\right)^{2}\right\}^{\frac{1}{2}}
\end{array}
\end{equation}

If all two conditions (3.25) and (3.26) are satisfied
simultaneously, the bound states exist. From Eq.(3.25) is seen that
if $l=0$, then one gets $n<0$. Hence, the Klein-Gordon equation for
the standard Woods-Saxon potential with zero angular momentum has no
bound states. For larger values of $V_{0}$ ($V_{0}>\frac{4\hbar
ac\sqrt{3l(l+1)}}{R_{0}^{2}}$) the condition (3.25) is not
satisfied. Therefore, no bound states exist for these values of
$V_{0}$.

According to Eq.(3.28) the energy eigenvalues  depend on the depth
of the potential $V_{0}$, the width of the potential $R_{0}$, and
the surface thickness $a$. Any energy eigenvalue must be less than
$V_{0}$. If constraints imposed on $n$ and $V_{0}$ are satisfied,
the bound states appear. From Eq.(3.26) is seen that the potential
depth increases when the parameter $a$ increases, but the parameter
$R_{0}$ is decreasing for given $l$ quantum number and vice versa.
Therefore, one can say that the bound states exist within this
potential. Thus, the energy spectrum Eq.(3.28) are limited, i.e., we
have only the finite number of energy eigenvalues.

The binding energy of a bound Klein-Gordon particle is defined as
[18]:
\begin{equation}
 E_{b}=E_{nl}-m_{0}c^{2}.
\end{equation}
By using the empirical values $r_{0}=1.285fm$ and $a=0.65 fm$
taken from Ref.[19] the potential depth $V_{0}=(40.5+0.13A)MeV$
and the radius of the nucleus $R_{0}=r_{0}A^{1/3}fm$ are
calculated for the atomic mass number of target nucleus
$A=40;\,\,56;\,\,66;\,\,92;\,\, 140;\,\,208$ and pions with mass
$m_{0}c^{2}=139.570 MeV$. In Table 1, energies of the bound states
obtained numerically for the spherical standard Woods-Saxon
potential for some values of $l$ and $n$ are given. From Table 1
is seen that for fixed $n$ the energy of the bound states
increases with increase of $l$. This means that due to the
centrifugal potential $V_{l}(r)$ in the system the repulsive
forces appear. Therefore, in order to compensate this potential
the energy of the bound state must increase [20].

In addition, we have seen that there are some restrictions on the
potential parameters in order to obtain bound state solutions. We
also point out that the exact results obtained for the standard
Woods-Saxon potential may have some interesting applications for
studying different quantum mechanical and nuclear scattering
problems.

However, in the nonrelativistic limit with the mapping
$E^{(R)}_{nl}-m_{0}c^{2}\rightarrow E^{(NR)}_{nl}$, according to
Appendix A for bound state energy eigenvalues, we obtain [8]:
\begin{equation}
\begin{array}{c}
E_{nl}=\frac{\hbar^{2}l(l+1)}{2m_{0}
R_{0}^2}\left(1+\frac{12a^{2}}{R_{0}^2}\right)-   \\
\frac{\hbar^{2}}{2m_{0}
a^{2}}\left[\frac{\left(\sqrt{1+\frac{192l(l+1)a^{4}}{R_{0}^4}}-2n-1\right)^{2}}{16}+
\frac{4\left(\frac{m_{0}V_{0}a^{2}}{\hbar^{2}}
-\frac{4l(l+1)a^{3}}{R_{0}^3}\right)^2}{\left(\sqrt{1+\frac{192l(l+1)a^{4}}{R_{0}^4}}-2n-1\right)^{2}}+\frac{m_{0}
V_{0}a^{2}}{\hbar^{2}}\right]
\end{array}
\end{equation}

Let us note, that in work [12] the Klein-Gordon and Schr\"odinger
equations with Woods-Saxon potential for $l\neq0$ states was solved,
the bound state energy eigenvalues and the corresponding
eigenfunctions were found. But in the work [12] errors were made in
application of the NU method to Schr\"odinger equation, which led,
as in papers [15, 16] to wrong predictions for the bound energy
eigenvalues and the corresponding eigenfunctions. Such errors were
made also in the works [9, 13, 14].

 Now, we are going to determine the
radial eigenfunctions of this potential. Having substituted $\pi(z)$
and $\sigma(z)$ into Eq.(2.4) and then solving first-order
differential equation, one can find the finite function $\Phi(z)$ in
the interval $(0,1)$

\begin{equation}
\Phi(z)=z^{\epsilon}\left(1-z\right)^{\sqrt{\epsilon^{2}-\beta^{2}+\gamma^{2}}}.
\end{equation}

It is easy to find the second part of the wave function from the
definition of weight function:
\begin{equation}
\rho(z)=z^{2\epsilon}\left(1-z\right)^{2\sqrt{\epsilon^{2}-\beta^{2}+\gamma^{2}}},
\end{equation}
and substituting into Rodrigues relation (2.5), we get
\begin{equation}
y_{n}(z)=B_{n}z^{-2\epsilon}\left(1-z\right)^{-2\sqrt{\epsilon^{2}-\beta^{2}+\gamma^{2}}}
\frac{d^{n}}{dz^{n}}\left[z^{n+2\epsilon}\left(1-z\right)^{n+2\sqrt{\epsilon^{2}-\beta^{2}+\gamma^{2}}}\right].
\end{equation}
where $B_{n}=\frac{1}{n!}$ is the normalization constant [21].
Then, $y_{n}$ is given by the Jacobi polynomials
\begin{equation}
y_{n}(z)=P_{n}^{\left(2\epsilon,2\sqrt{\epsilon^{2}-\beta^{2}+\gamma^{2}}
\right)}(1-2z),
\end{equation}
where $$P_{n}^{(\alpha,\beta)}(1-2z)=\frac{1}{n!}
z^{-\alpha}\left(1-z\right)^{-\beta}\frac{d^{n}}{dz^{n}}
\left[z^{n+\alpha}\left(1-z\right)^{n+\beta}\right].$$

The corresponding $u_{nl}(z)$ radial wave functions are found to be
\begin{equation}
u_{nl}(z)=C_{nl}z^{\epsilon}\left(1-z\right)^{\sqrt{\epsilon^{2}-\beta^{2}+\gamma^{2}}}P_{n}^{\left
(2\epsilon,\,2\sqrt{\epsilon^{2}-\beta^{2}+\gamma^{2}}\right)}(1-2z),
\end{equation}
where $C_{nl}$ is the normalization constants determined using
$\int_o^\infty[u_{nl}(r)]^2dr=1$ constraint, i.e.,
$$aC^{2}_{nl}\int_o^1z^{2\epsilon-1}\left(1-z\right)^{2\sqrt{\epsilon^{2}-\beta^{2}+\gamma^{2}}-1}
\left[P^{\left(2\varepsilon,
2\sqrt{\epsilon^{2}-\beta^{2}+\gamma^{2}}\right)}_{n}\left(1-2z\right)\right]^{2}dz=1.$$

\section{\bf Conclusion}

In this paper, we have analytically calculated energy eigenvalues of
the bound states and corresponding eigenfunctions in the new exactly
solvable Woods-Saxon potential. The energy eigenvalue expression for
Woods-Saxon potentials is given by Eq.(3.28). The nonrelativistic
limit (3.30) of the bound state energy spectrum was also obtained .
As it should be expected, for any given set of parameters $V_{0},
R_{0}$ and $a$, the energy levels of standard Woods-Saxon potential
are positive. The obtained results are interesting for both
theoretical and experimental physicists, because they provide exact
expression for energy eigenvalues and corresponding eigenfunctions.

\newpage

\appendix
\section{}
\label{AppendixA}

After some simple transformations  Eq.(3.27) takes the form:
\begin{equation}
\begin{array}{c}
E^{(R)}_{nl}=-\frac{V_{0}}{2}\left(1-\frac{l(l+1)(C_{1}+C_{2})}{\alpha
^{2}\left(n'^{2}+\frac{V_{0}^{2}a^{2}}{\hbar
^{2}c^{2}}\right)}\right)+m_{0}c^{2}\left\{1+\frac{\hbar
^{2}l(l+1)C_{0}}{m_{0}^{2}c^{2}R_{0}^{2}}-\frac{V_{0}^{2}a^{2}}{\hbar
^{2}c^{2}\left(n'^{2}+\frac{V_{0}^{2}a^{2}}{\hbar
^{2}c^{2}}\right)}- \right.   \\  \left.\frac{4\hbar
^{2}l(l+1)C_{0}V_{0}^{2}a^{2}}{\hbar^{2}c^{4}m_{0}^{2}R_{0}^{2}
\left(n'^{2}+\frac{4V_{0}^{2}a^{2}}{\hbar^{2}c^{2}}\right)}-\frac{\hbar
^{2}n'^{2}}{4m_{0}^{2}c^{2}a^{2}}\left(1-\frac{l(l+1)(C_{1}+C_{2})}{\alpha
^{2}\left(n'^{2}+\frac{4V_{0}^{2}a^{2}}{\hbar
^{2}c^{2}}\right)}\right)^2\right\}^{\frac{1}{2}}
\end{array}
\end{equation}

Expanding, finally, the above eigenvalue in a series of powers of
$\frac{1}{c^{2}}$ yields:

\begin{equation}
\begin{array}{c}
E^{(R)}_{nl}=-\frac{V_{0}}{2}\left[1-\frac{4l(l+1)(C_{1}+C_{2})}
{\alpha^{2}\left(\sqrt{1+\frac{192l(l+1)a^{4}}{R_{0}^{4}}}-2n-1
\right)^2}\right]+m_{0}c^{2}\left\{1+\frac{\hbar^{2}l(l+1)C_{0}}
{2m_{0}^{2}c^{2}R_{0}^{2}}-\right.  \\   \left.
\frac{4V_{0}^{2}a^{2}}{2\hbar^{2}c^{2}
\left(\sqrt{1+\frac{192l(l+1)a^{4}}{R_{0}^{4}}}-2n-1\right)^{2}}-
\frac{\hbar^{2}}{32m_{0}^{2}c^{2}a^{2}}
\left(\sqrt{1+\frac{192l(l+1)a^{4}}{R_{0}^{4}}}-2n-1\right)^{2}
\times \right.   \\   \left.     \left[1-
\frac{4l(l+1)(C_{1}+C_{2})}{\alpha^{2}
\left(\sqrt{1+\frac{192l(l+1)a^{4}}{R_{0}^{4}}}-2n-1\right)^{2}}\right]^{2}\right\}+
\dots=-\frac{V_{0}}{2}+ \frac{2l(l+1)(C_{1}+C_{2})V_{0}}{\alpha^{2}
\left(\sqrt{1+\frac{192l(l+1)a^{4}}{R_{0}^{4}}}-2n-1\right)^{2}}-   \\
m_{0}c^{2}+\frac{\hbar^{2}l(l+1)C_{0}}{2m_{0}R_{0}^{2}}-
\frac{2m_{0}V_{0}^{2}a^{2}}{\hbar^{2}
\left(\sqrt{1+\frac{192l(l+1)a^{4}}{R_{0}^{4}}}-2n-1\right)^{2}}-
\frac{\hbar^{2}}{32m_{0}a^{2}}\left(\sqrt{1+\frac{192l(l+1)a^{4}}{R_{0}^{4}}}-2n-1\right)^{2}+   \\
\frac{\hbar^{2}l(l+1)(C_{1}+C_{2})}{4m_{0}a^{2}\alpha^{2}}-
\frac{\hbar^{2}l^{2}(l+1)^{2}(C_{1}+C_{2})^{2}}{2m_{0}a^{2}\alpha^{4}
\left(\sqrt{1+\frac{192l(l+1)a^{4}}{R_{0}^{4}}}-2n-1\right)^{2}}+\dots=
m_{0}c^{2}+\frac{\hbar^{2}l(l+1)C_{0}}{2m_{0}R_{0}^{2}}-\frac{V_{0}}{2}  \\
- \frac{1}{16}\frac{\hbar^{2}}{2m_{0}a^{2}}
\left(\sqrt{1+\frac{192l(l+1)a^{4}}{R_{0}^{4}}}-2n-1\right)^{2}-
\frac{\hbar^{2}}{2m_{0}a^{2}}\frac{\left(\frac{2m_{0}V_{0}a^{2}}{\hbar^{2}}-
\frac{l(l+1)(C_{1}+C_{2})}{\alpha^{2}}\right)^{2}}{\left(\sqrt{1+
\frac{192l(l+1)a^{4}}{R_{0}^{4}}}-2n-1\right)^{2}}+  \\
\frac{\hbar^{2}l(l+1)(C_{1}+C_{2})}{4m_{0}a^{2}\alpha^{2}}+\dots=
m_{0}c^{2}+\frac{\hbar^{2}l(l+1)C_{0}}{2m_{0}R_{0}^{2}}-
\frac{\hbar^{2}}{2m_{0}a^{2}}\left[\frac{1}{16}
\left(\sqrt{1+\frac{192l(l+1)a^{4}}{R_{0}^{4}}}-2n-1\right)^{2}+\right.
\\  \left.  \frac{1}{2}\left(\frac{2m_{0}V_{0}a^{2}}{\hbar^{2}}-
\frac{l(l+1)(C_{1}+C_{2})}{\alpha^{2}}\right)+
\frac{\left(\frac{2m_{0}V_{0}a^{2}}{\hbar^{2}}-
\frac{l(l+1)(C_{1}+C_{2})}{\alpha^{2}}\right)^{2}}
{\left(\sqrt{1+\frac{192l(l+1)a^{4}}{R_{0}^{4}}}-2n-1\right)^{2}}
\right]+\dots=m_{0}c^{2}+   \\
\frac{\hbar^{2}l(l+1)C_{0}}{2m_{0}R_{0}^{2}}-
\frac{\hbar^{2}}{2m_{0}a^{2}}\left[\frac{1}{4}
\left(\sqrt{1+\frac{192l(l+1)a^{4}}{R_{0}^{4}}}-2n-1\right)+
\frac{\frac{2m_{0}V_{0}a^{2}}{\hbar^{2}}-\frac{l(l+1)(C_{1}+C_{2})}
{\alpha^{2}}}{\sqrt{1+\frac{192l(l+1)a^{4}}{R_{0}^{4}}}-2n-1}
\right]^{2}+o\left(\frac{1}{c^{4}}\right)
\end{array}
\end{equation}

Denoting, as usual, the energy eigenvalues in the nonrelativistic
case by $E^{(R)}_{nl}-m_{0}c^{2} \rightarrow E^{(NR)}_{nl}$, we
have:

\begin{equation}
\begin{array}{c}
E^{(NR)}_{nl}=\frac{\hbar^{2}l(l+1)C_{0}}{2m_{0}R_{0}^{2}}-
\frac{\hbar^{2}}{2m_{0}a^{2}}\left[\frac{1}{4}
\left(\sqrt{1+\frac{192l(l+1)a^{4}}{R_{0}^{4}}}-2n-1\right)+
\frac{\frac{2m_{0}V_{0}a^{2}}{\hbar^{2}}-\frac{l(l+1)(C_{1}+C_{2})}
{\alpha^{2}}}{\sqrt{1+\frac{192l(l+1)a^{4}}{R_{0}^{4}}}-2n-1}
\right]^{2}
\end{array}
\end{equation}
Thus, substituting the expressions of $C_{0}, C_{1}$ and $C_{2}$
into the above energy eigenvalues $E^{(NR)}_{nl}$, we obtain:

\begin{equation}
\begin{array}{c}
E^{(NR)}_{nl}=\frac{\hbar^{2}l(l+1)}{2m_{0}R_{0}^{2}}
\left(1-\frac{4a}{R_{0}}+\frac{12a^{2}}{R_{0}^{2}}\right)-   \\
\frac{\hbar^{2}}{2m_{0}a^{2}}\left[\frac{1}{4}
\left(\sqrt{1+\frac{192l(l+1)a^{4}}{R_{0}^{4}}}-2n-1\right)+
\frac{2\left(\frac{m_{0}V_{0}a^{2}}{\hbar^{2}}-\frac{4l(l+1)a^{3}}
{R_{0}^{3}}\right)}{\sqrt{1+\frac{192l(l+1)a^{4}}{R_{0}^{4}}}-2n-1}\right]^{2}
\end{array}
\end{equation}

or

\begin{equation}
\begin{array}{c}
E^{(NR)}_{nl}=\frac{\hbar^{2}l(l+1)}{2m_{0}
R_{0}^{2}}\left(1+\frac{12a^{2}}{R_{0}^{2}}\right)-    \\
\frac{\hbar^{2}}{2m_{0}a^{2}}\left[\frac{\left(\sqrt{1+\frac{192l(l+1)a^{4}}{R_{0}^{4}}}-2n-1\right)^{2}}{16}+
\frac{4\left(\frac{m_{0}V_{0}a^{2}}{\hbar^{2}}-\frac{4l(l+1)a^{3}}{R_{0}^{3}}\right)^{2}}
{\left(\sqrt{1+\frac{192l(l+1)a^{4}}{R_{0}^{4}}}-2n-1\right)^{2}}+\frac{m_{0}V_{0}a^{2}}{\hbar^{2}}\right]
\end{array}
\end{equation}

\newpage
\begin{table}[h]
\begin{center}
\begin{tabular}{|c|c|c|c|c|c|}\hline
$A$ & $R_{0}, fm$ & $V_{0}, MeV$ & $n$ & $l$ & $E_{b}, MeV$ \\
\hline
    40  &  4.3946  &   45.70    &  0  &  1  &   -107.8777   \\ \hline
    56  &  4.9162  &   47.78    &  0  &  1  &   -127.5238   \\ \hline
    56  &  4.9162  &   47.78    &  0  &  2  &    -17.5985   \\ \hline
    66  &  5.1930  &   49.08    &  0  &  2  &    -50.3359   \\ \hline
    92  &  5.8010  &   52.46    &  0  &  2  &   -101.8967   \\ \hline
   140  &  6.6724  &   58.70    &  0  &  3  &    -92.5327   \\ \hline
   208  &  7.6136  &   67.54    &  0  &  4  &   -105.0865   \\ \hline
   208  &  7.6136  &   67.54    &  0  &  5  &    -33.6014   \\ \hline
\end{tabular}
\end{center}

\caption{Energies of the bound states for the Woods-Saxon
potential for different values of $n,\,\, l$ calculated using
Eqs.(3.25), (3.26), (3.28) and (3.29).} \label{table1}
\end{table}


\newpage

\begin{thebibliography}{99}
\bibitem{1} F. Cooper, A. Khare  and  U. Sukhatme,  Phys. Rep. {\bf 251} (1995) 267.
\bibitem{2} D. A. Morales,  Chem. Phys. Lett. {\bf 394} (2004) 68.
\bibitem{3} C. L. Pekeris, Phys. Rev. {\bf 45} (1934) 98.
\bibitem{4} S. Fl\"{u}gge, Practical Quantum Mechanics, Vol.1 (Springer, Berlin, 1994)
\bibitem{5} O. Bayrak  and I. Boztosun,  J. Phys. A {\bf 39} (2006) 6955 (arxiv: nucl-th / 0604042 V1).
\bibitem{6} O. Bayrak, G. Kocak  and I. Boztosun, J. Phys. A: Math. Gen. {\bf 39} (2006) 11521 (arxiv: math-ph / 0609010 V1).
\bibitem{7} V. H. Badalov, H. I. Ahmadov and S. V. Badalov,  News of Baku University, N{\bf 2} (2008) 157.
\bibitem{8} V. H. Badalov, H. I. Ahmadov and A. I. Ahmadov,  Int.J.Mod.Phys. E {\bf 18} (2009) 631 (arxiv: math-ph / 0905.273 V1).
\bibitem{9} S. M. Ikhdair  and  R. Sever, (arxiv: quant-ph /0610183 V1).
\bibitem{10} H. Egrifes and R. Sever, Int. J. Theor. Phys. {\bf 46} (2007) 935 (arxiv: quant-ph / 0609231).
\bibitem{11} R. D. Woods  and D. S. Saxon ,  Phys. Rev.  {\bf 95} (1954) 577.
\bibitem{12} A. Arda  and  R. Sever, Int. J. Mod. Phys. C {\bf 20} (2009) 651 (arxiv: math-ph /0901.2773 V1).
\bibitem{13} A. Arda  and  R. Sever, Int.J.Mod.Phys. A {\bf 24} (2009) 3985 (arxiv: quant-ph/0902.2088 V1).
\bibitem{14} A.Berkdemir, C.Berkdemir and R.Sever, Mod.Phys.Lett. A {\bf 21}, 2087 (2006) (arxiv: quant-ph/0410153 V3).
\bibitem{15} C. Berkdemir, A. Berkdemir and R. Sever, Phys. Rev. C {\bf 72} (2005) 027001.
\bibitem{16} Editorial Note to Phys. Rev. C {\bf 72}, (2005) 027001; Phys. Rev. C {\bf 74} ( 2006) 039902(E).
\bibitem{17} A. F. Nikiforov  and V. B. Uvarov, Special Functions of Mathematical Physics (Birkh\"{a}user, Basel, 1988).
\bibitem{18} W. Greiner, Relativictic Quantum Mechanics (Springer, Berlin, 1990).
\bibitem{19} C. M. Perey, F. G. Perey, J. K. Dickens  and R. J. Silva, Phys. Rev. {\bf 175} (1968) 1460.
\bibitem{20} L.E. Ballentine, Quantum Mechanics (World Scientific Publishing, Singapore, 1998).
\bibitem{21} H. Bateman  and A. Erdelyi,  Higher Transcendental functions, Vol.2 (McGraw-Hill, New York, 1953)



\end{thebibliography}
\end{document}